\documentclass[12pt]{article}
\usepackage{amsfonts, graphicx,caption,subfig,bbm}

\newcommand{\C}{{\mathbb{C}}}
\newcommand{\Z}{{\mathbb{Z}}}

\def\be{\begin{equation}}
\def\ee{\end{equation}}
\def\ben{\begin{displaymath}}
\def\een{\end{displaymath}}
\def\ba{\begin{array}{c}}
\def\ea{\end{array}}

\newcommand{\ed}{\end{document}}

\begin{document}

\titlepage
\vspace*{2cm}

\begin{center}{\Large \bf

Scattering in the ${\cal PT}$-symmetric Coulomb potential

}\end{center}

\vspace{10mm}

\begin{center}

G\'{e}za L\'{e}vai

\vspace{3mm}

Institute of Nuclear Research of the Hungarian Academy of Sciences (ATOMKI),

PO Box 51, H-4001 Debrecen, Hungary,
\vspace{3mm}

and \vspace{3mm}

Petr Siegl \vspace{3mm}

Nuclear Physics Institute of Academy of Sciences of the Czech
Republic, 25068 \v{R}e\v{z}, Czech Republic,

Faculty of Nuclear Sciences and Physical Engineering, Czech Technical University, 11519 Prague, Czech Republic,

Laboratoire Astroparticules et Cosmologie, Universit\'e Paris Diderot, 75205 Paris, France
\vspace{3mm}

and \vspace{3mm}

Miloslav Znojil \vspace{3mm}

Nuclear Physics Institute of Academy of Sciences of the Czech
Republic, 25068 \v{R}e\v{z}, Czech Republic

\vspace{3mm}

\end{center}

\vspace{5mm}

\vspace{5mm}
\newpage

\section*{Abstract}
Scattering on the ${\cal PT}$-symmetric Coulomb potential is studied 
along a U-shaped trajectory circumventing the origin in the complex 
$x$ plane from below. This trajectory reflects ${\cal PT}$ symmetry,  
sets the appropriate boundary conditions for bound states and also 
allows the restoration of the correct sign of the energy eigenvalues. 
Scattering states are composed from the 
two linearly independent solutions valid for non-integer values of the $2L$ 
parameter, which would correspond to the angular momentum in 
the usual Hermitian setting. Transmission and reflection coefficients 
are written in closed analytic form and it is shown that similarly 
to other ${\cal PT}$-symmetric scattering systems the latter exhibit 
handedness effect. Bound-state energies are recovered from the 
poles of the transmission coefficients.

\newpage

\section{Introduction}

The Kepler--Coulomb problem has always played a special role in the 
formulation and the application of quantum mechanics. Besides being  
one of the textbook examples for exactly solvable problems, it also 
exhibits features that have always attracted the attention of mathematical 
physicists. Among these one can mention that the Coulomb potential 
possesses both discrete and continuous spectra, which can be associated 
with dynamical symmetry and Lie algebras describing them (see e.g. 
Ref. \cite{wyb} for a review). Although in the description of 
realistic physical systems the three-dimensional Coulomb potential 
and the associated radial Schr\"odinger equation is used in most 
cases, much work has been done extending the Coulomb potential to 
other dimensions. Of these the one-dimensional Coulomb potential 
is the most notable, as the singularity at $x=0$ raises interesting 
questions both for the $V(x)\sim -x^{-1}$ and the 
$V(x)\sim -\vert x\vert^{-1}$ potentials. (See e.g. Ref. \cite{ose93} 
and references.) The discussion of this seemingly humble system 
requires techniques like the self-adjoint extension of the relevant 
differential operator \cite{fis95}. 

Manifestly non-Hermitian versions of the Coulomb potential have also
been studied in terms of ${\cal PT}$-symmetric quantum mechanics.
In this theory \cite{bb98} Hamiltonians invariant with respect to 
the simultaneous ${\cal P}$ space and ${\cal T}$ time inversion 
were found to exhibit features characteristic for Hermitian 
systems, such as partly or fully real energy spectrum and the 
conservation of the norm. (See e.g. \cite{bender07} for a recent review.) 

The first examples for ${\cal PT}$-symmetric potentials were of 
the type $V(x)=x^2({\rm i}x)^{\epsilon}$, including 
the archetypal imaginary cubic potential for $\epsilon=1$. After the 
first numerical results, the conjecture of the reality of the energy 
spectrum was proven analytically for such potentials \cite{DDT}. 
An interesting aspect of these systems is that often they cannot be 
defined on the real $x$ coordinate axis, rather their solutions 
are normalizable only along certain trajectories of the complex 
$x$ plane. This was the case, for example, for the above potential with 
$\epsilon\ge 2$, when these trajectories had to fall into wedges 
lying in the lower half of the plane in $\epsilon$-dependent positions 
symmetrical with respect to the imaginary axis. 

Later the ${\cal PT}$-symmetric version of a number of exactly solvable 
potentials have been formulated, mainly along the real $x$ axis 
or along a line parallel with it $x-{\rm i}c$. (See e.g. Ref. 
\cite{jpa00} and references.) The importance of this imaginary shift was 
that singularities lying on the real $x$ axis could be avoided, 
and at the same time, the energy spectrum remained independent of $c$. 
In this way real potentials defined on the positive half axis could 
be extended to negative $x$ values too. As the result of this, 
solutions irregular in the Hermitian case became regular in the 
${\cal PT}$-symmetric version of the potential, and this led to 
a richer energy spectrum. 

The Coulomb potential was among the first exactly solvable potentials 
considered within the ${\cal PT}$-symmetric setting. It was found, 
however, that it cannot be defined on the real $x$ axis because the 
solutions were not regular for both $x\rightarrow \infty$ and 
$x\rightarrow -\infty$ \cite{jpa00}. In Ref. \cite{pla00} a parabolic 
trajectory was proposed, which was inspired by the ${\cal PT}$-symmetrized 
version of the well-known harmonic oscillator--Coulomb mapping. 
This study revealed that the spectrum of the ${\cal PT}$-symmetric 
Coulomb potential includes a second set of discrete energy 
eigenvalues in addition to the one that is present also in the 
spectrum of the real Coulomb potential. But as a more interesting 
feature, the energy spectrum was inverted \cite{pla00}. The  
interpretation of this unusual finding was given later in 
Ref. \cite{pla09}. The transformation properties of the solutions under the 
${\cal PT}$ operation including solutions both with real 
and with pairwise complex conjugate energy eigenvalues 
were also discussed \cite{GLPR}. Another study of the ${\cal PT}$-symmetric 
Coulomb potential was done in Ref. \cite{asrr02}: there the 
Coulomb potential was defined as $V(x)\sim\vert x-{\rm i}c\vert^{-1}$. 

In the present work we extend,  to the scattering scenario,
the discussion of the ${\cal
PT}$-symmetric Coulombic bound states as presented in refs.
\cite{pla00,pla09}. In this direction we
intend to pursue two ideas.
The first one reflects the existence of a number of  publications
\cite{jpa01,jpa02b,ahmedxx,glem09,fcjpdav}
where the standard ${\cal PT}$-symmetric version of the
scattering problem has already been described and developed in
application to a number
of exactly solvable potentials. We feel
inspired by the observation that in all of these works the
scattering has only been considered along the real $x$ axis
and/or along a trivially complexified, shifted straight-line
contour $x(s)=s-{\rm i}c$ with real variable $s$ and constant $c$.
We shall change this perspective by employing an
utterly nontrivial negative-mass
generalization of the complex integration path
$x=x_{(\varepsilon)}^U(s)$ as
already proposed, in the context of the stabilization of
the Coulomb bound
states, in ref. \cite{pla09} (see also eq. (\ref{urva})
below).
In this setting we shall reveal the new role of the real
parameter $\varepsilon$ which appears to bring a new degree of
freedom in the
phenomenological scattering theory, indeed.

The second motivation of our present interest in the ${\cal
PT}$-symmetric Coulombic scattering
along a U-shaped contour $x(s)$ of ref. \cite{pla09}
is more physical since it
reflects the unique possibility of the {\em coexistence} of
discrete and scattering states in a single potential
(in this respect cf., e.g., the review  \cite{wyb} once more).
We have to
emphasize that
in the ${\cal PT}$-symmetric context
such a feature
has not yet been achieved even in the
models of scattering along curved complex contours
(cf., e.g., Ref. \cite{toboggan}
where the ``tobogganic" integration
path $x=x(s)$ has been chosen even as extending, in principle, along
several Riemann sheets of the scattering wavefunctions $\psi\left(x(s)\right)\,$).
Thus, the present U-shaped choice of $x^U_{(\varepsilon)}(s)$
will represent
physics which varies with the ``contour width" $\varepsilon$. This
seems to offer a scattering-scenario analogue
of the variability of
the bound-state spectra mediated, according to Refs.
\cite{bb98,toboggan}, by the variability of our choice of
the Stokes' ``wedges" in the
complex $x$ plane.

The structure of the paper is as follows. In Section \ref{defi} the 
general formulation of the problem is presented, together with the 
U-shaped trajectory along which the scattering problem is considered. 
Section \ref{cscatt} deals with the actual calculation of the 
transmission and reflection coefficients, while the results are 
summarized in Section \ref{sum}.

\section{Definition of the problem}
\label{defi}

Let us consider the Schr\"odinger equation 
 \be
 \frac{\ \hbar^2}{2m} \,
  \left [-\frac{{\rm d}^2}{{\rm d}{x}^2}
  +  \frac{L(L+1)}{{x}^2}\right ]\, \Psi({x})
 +V(x) \, \Psi({x}) = E \,\Psi({x})\,,
 \label{SEor}
  \ee
defined in the $x$ variable, which runs along a trajectory of the 
complex $x$ plane. Let us assume that this trajectory can be parametrized 
in terms of a real variable as $x(s)$. In order to implement ${\cal PT}$ 
symmetry of this system we introduce 
 \be
 V(x)=\frac{{\rm i}Z}{x}\, ,
 \label{coulovani}
 \ee
where $Z$ and $L(L+1)$ are real. This latter condition is met if 
$L$ is chosen real. The two linearly independent solution of (\ref{SEor}) 
can be written in terms of confluent hypergeometric functions as 
 \be
 \Psi_1(x)=C_1 \,e^{-kx}x^{L+1} 
 {_1}{F}{_1}(1+L+{{\rm i}Z}/{(2k)},2L+2,2kx)
  \label{gsol1}
 \ee
 \be
 \Psi_2(x)=C_2\,e^{-kx}x^{-L} 
 {_1}{F}{_1}(-L+{{\rm i}Z}/{(2k)},-2L,2kx)\ ,
  \label{gsol2}
 \ee
where $2mE/\hbar^2=-k^2$. Note that (\ref{gsol1}) and (\ref{gsol2}) 
represent 
the two linearly independent solutions only if $2L\ \notin \Z $.  

Equation (\ref{SEor}) with (\ref{coulovani}) differs from the usual 
radial Schr\"odinger equation of the Coulomb problem in the complexified 
potential, and also in the trajectory it is defined on. In a systematic 
reformulation of real solvable potentials and the respective bound states to 
their ${\cal PT}$-symmetric counterpart it was found \cite{jpa00} that 
this problem cannot be defined on the real $x$ axis or its imaginary 
shifted version $x-{\rm i}c$, because the boundary conditions cannot 
be implemented in both directions due to the exponential factor in 
(\ref{gsol1}) and (\ref{gsol2}). In an effort to determine the 
genuine ${\cal PT}$-symmetric version of the Coulomb potential, the 
well-known Coulomb--harmonic oscillator mapping was used: this  
transformation was applied to the ${\cal PT}$-symmetric harmonic 
oscillator defined on the imaginary shifted real axis $x-{\rm i}c$ 
\cite{pla00}. The resulting trajectory was a parabola in the first and 
fourth quadrant, 
circumventing the origin from the left. In order to make it 
${\cal PT}$-symmetric, i.e. left-right symmetric in the coordinate 
space, it had to be tilted to the first and second quadrant by 
the multiplication ${\rm i}x$. In order to keep the $kx$ quantity 
intact, $k$ also had to be tilted in the opposite direction in the $k$ 
wave number space as $-{\rm i}k$. This resulted 
in the unusual finding that the energy spectrum was inverted, as can 
be seen from the relation $2mE/\hbar^2=-k^2$. 
Note that the two sets of discrete-energy solutions discussed in  
\cite{pla00} are obtained from (\ref{gsol1}) and 
(\ref{gsol2}) by 
substituting a non-positive integer in the first argument of the 
respective confluent hypergeometric functions, reducing them to 
the expected generalized Laguerre polynomial form \cite{as70}. 
(In order to match the formulae of Ref. \cite{pla00} with those obtained 
here the 
following substitutions have to be made: $t\rightarrow x$, 
$A\rightarrow L+1/2$, $e^2=1$, while $\kappa^2$ should be 
chosen $Z/[2(n+L+1)]$ and $Z/[2(n-L)]$ in the two cases, corresponding 
to the two possible values of the quasi-parity $q=\pm 1$.) 

As another possible trajectory, a U-shaped curve circumventing 
the origin, illustrated in the Fig.\ref{Ucurve}, was proposed in Ref. \cite{pla09}. It is defined for 
a suitable $\varepsilon>0$ as 
 \be
 x(s)=x^{U}_{(\varepsilon)}(s)\,= \,
 \left \{
 \begin{array}{ll}
 -{\rm i}(s+\frac{\pi}{2}\varepsilon)
 -\varepsilon, & s \in (-\infty, -\frac{\pi}{2}\varepsilon),\\
 \varepsilon e^{ {\rm i}({s/\varepsilon+3/2\pi }
  )},&  s \in
 ( -\frac{\pi}{2}\varepsilon,
 \frac{\pi}{2}\varepsilon),\\
  {\rm i}(s-\frac{\pi}{2}\varepsilon)+\varepsilon\,, & s \in
(\frac{\pi}{2}\varepsilon, \infty).
 \ea
 \right .
 \label{urva}
 \ee
\begin{figure}[ht]
	\centering
		\includegraphics[width=0.50\textwidth]{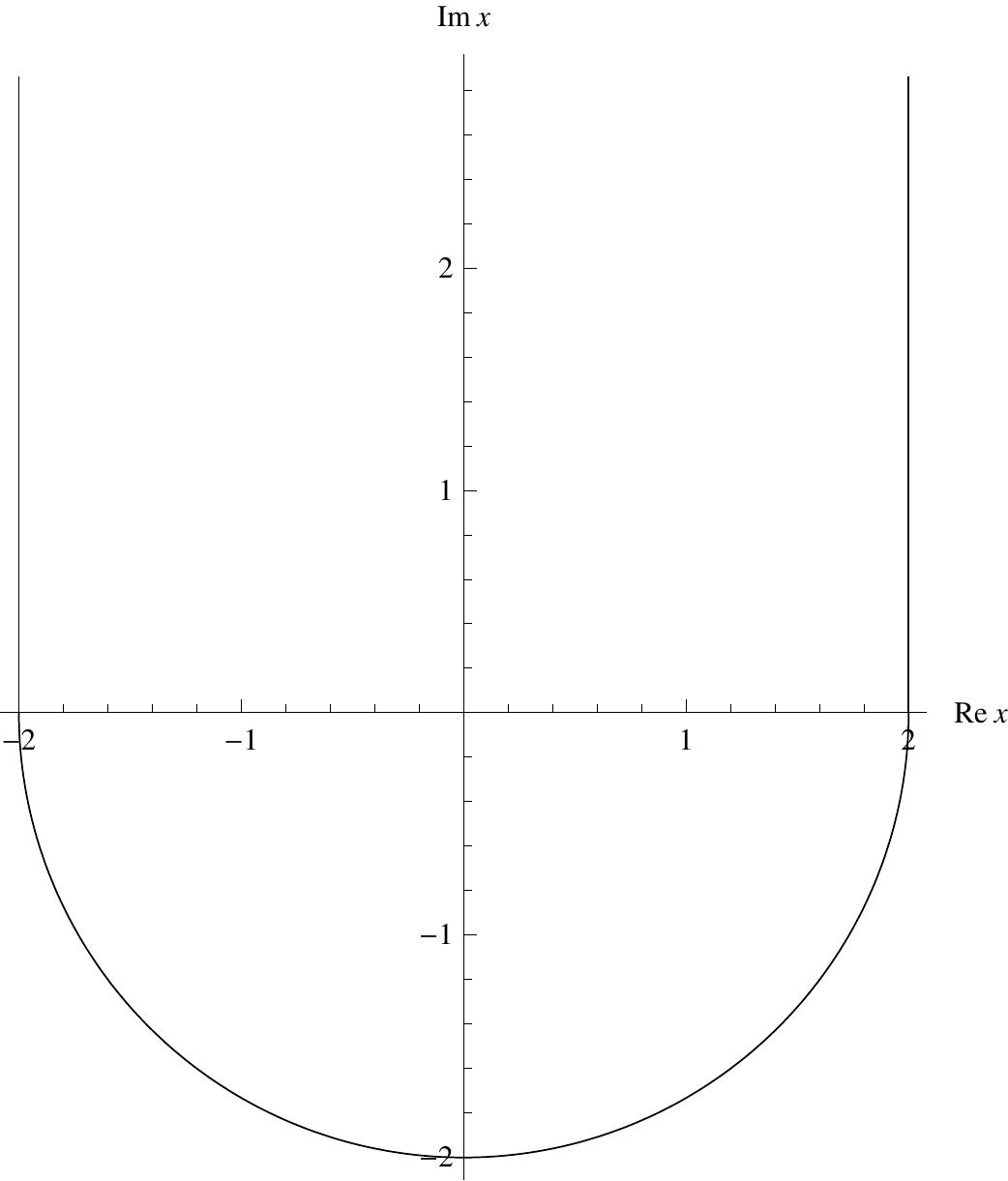}
	\caption{U-shaped curve in a complex $x$ plane for $\varepsilon=2$}
	\label{Ucurve}
\end{figure}
The asymptotic $\varepsilon-$dependence of this curve has an
immediate physical meaning because it enables us to distinguish
between non-equivalent alternative asymptotes of the curves
$x(s)$ along which the non-equivalent asymptotic boundary
conditions will be specified for our wave functions. Of course,
in contrast to the scattering wave functions which must be
different on the left and right asymptotic branch of $x(s)$, all
of the curves of coordinates exhibit the same left-right
symmetry $x(-s) = -x^*(s)$ 
in the complex $x$ plane, which combines spatial reflection ${\cal P}$ 
with complex conjugation ${\cal T}$ that mimics time-reversal. 
It is seen that for large $\vert s\vert$ (in fact, for
$\vert s\vert \gg \varepsilon\pi/2$) the solutions behave as
$\exp(\pm {\rm i}k\vert s\vert)$. For real $k=(-2mE)^{1/2}/\hbar$
(i.e., for $m>0$ and $E<0$ of Ref. \cite{pla00} or for $m<0$ and
$E>0$
in \cite{pla09}), this
represents an oscillatory solution. In parallel, for an imaginary $k$
(i.e., for $m>0$ and $E>0$ or  for $m<0$ and $E<0$) it
corresponds to exponentially decaying or growing solutions, depending
on the sign of ${\rm Im}(k)$.

The next result of this analysis presented also in Ref.
\cite{pla09}
is that at the negative mass,
the U-shaped parametrization (\ref{urva}) opens the way towards
the {\em simultaneous} description of the bound and scattering
states. In this setting the role of the asymptotic physical
coordinate is played by the real parameter $s$ of course.

The matching of logarithmic derivatives is usually used in models where
the potential $V$ is, at some point $x(s_0)$ of the curve of 
(possibly, complexified) coordinates, discontinuous. For analytic 
potentials (leading to analytic wave functions $\psi(x)$), the 
situation is different since these functions are usually well 
defined in all the points of some Riemann surface ${\cal S}$.  This 
means that in general, our analytic wave functions are multivalued 
functions which become single-valued, typically, on any selected 
Riemann sheet specified, say, as a subdomain ${\cal D}$ of a cut 
complex plane $\C$. In such a scenario it is only necessary to 
match the logarithmic derivatives of our analytic wave functions 
$\psi(x(s))$ during transition of the path $x(s)$ between 
neighboring Riemann sheets of the Riemann surface (i.e., typically,
between  pairs of non-overlapping subdomains ${\cal D}_{\pm}$ of  
$\C$). For this purpose it is usually sufficient to employ the 
analyticity of $\psi(x)$ and to simplify the matching via a suitable 
deformation of the path $x(s)$. Thus, most easily, one may analyse 
the transition between ${\cal D}_{+}$ and ${\cal D}_{-}$ just in an 
arbitrarily small vicinity of a branch point where  functions 
$\psi(x)$ degenerate to their dominant parts with trivial analytic-continuation properties.

\section{Scattering in the ${\cal PT}$-symmetric Coulomb potential}
\label{cscatt}

In what follows we shall make use of parametrization (\ref{urva}) to
study scattering on the ${\cal PT}$-symmetric Coulomb potential
at negative mass.
In Ref. \cite{pla09} this unusual option has been explained as
making the ${\cal PT}-$symmetric Coulomb bound states stable.
Here, we shall emphasize
that such an option is also necessary for a consistent description
of the scattering along the U-shaped complex contour.

On a purely technical level we shall employ
the
natural analytic continuity of functions (\ref{gsol1}) and
(\ref{gsol2}). In order to facilitate the implementation of
this idea the auxiliary complex phase factors
$\exp(2\pi{\rm i}(L+1))$ and $\exp(2\pi{\rm i}(-L))$
will  be introduced in the two solutions for Re$(x)>0$.
Then, the asymptotic expansion of the solutions can be written as \cite{Luke} 
\begin{eqnarray}
\begin{array}{lcl}
{_1}{F}{_1}(a,b,z)&\sim& \frac{\Gamma(b)}{\Gamma(b-a)}(z^{-1}
 {\rm e}^{{\rm i}\pi})^a {_2}{F}{_0}(a,1+a-b,-z^{-1})   +\\
                  &    & \frac{\Gamma(b)}{\Gamma(a)}e^z z^{a-b}
		  {_2}{F}{_0}(b-a,1-a,z^{-1})\ ,
\label{fass}
\end{array}
 \end{eqnarray}
where ${\rm Im}(z)>0$, $\vert {\rm Arg}(z)\vert <\pi$ as 
$\vert z\vert\longrightarrow \infty$. 

Applying (\ref{fass}) to (\ref{gsol1}) and (\ref{gsol2}) and employing 
the parametrization of $x$ (\ref{urva}) the following 
asymptotic expansions are obtained for $|s|\to \infty$: 
\begin{eqnarray}
\begin{array}{lcl}
\psi_{j}(s\rightarrow -\infty)  &\sim 
        & a_{j-}  {\rm e}^{{\rm i}(k s -\frac{Z}{2k}\ln(-2ks))} + 
          b_{j-}  {\rm e}^{-{\rm i}(k s -\frac{Z}{2k}\ln(-2ks))}
\label{psiassn}
 \\
\psi_{j}(s\rightarrow \infty)  &\sim 
        & a_{j+}  {\rm e}^{{\rm i}(k s +\frac{Z}{2k}\ln(2ks))} + 
          b_{j+}  {\rm e}^{-{\rm i}(k s +\frac{Z}{2k}\ln(2ks))}\ ,
\label{psiassp}
\end{array}
\end{eqnarray}
where $j=1,\ 2$. The logarithmic terms in the exponentials are 
characteristic of the Coulomb asymptotics and indicate that the 
Coulomb potential vanishes slower than genuine short-range potentials 
exhibiting exponential tail, for example \cite{newton}. 
The coefficients are 
\begin{eqnarray}
a_{1+}&=&C_1(2k)^{-L-1}{\rm e}^{-{\rm i}k\pi\varepsilon/2}
{\rm e}^{{\rm i}\pi(2L+2)}{\rm e}^{-\frac{\pi Z}{4k}}
{\rm e}^{k\varepsilon}\frac{\Gamma(2L+2)}{\Gamma(L+1+{\rm i}Z/(2k))}
\\
a_{1-}&=&C_1(2k)^{-L-1}{\rm e}^{{\rm i}k\pi\varepsilon/2}
{\rm e}^{{\rm i}\pi(L+1)}{\rm e}^{-\frac{\pi Z}{4k}}
{\rm e}^{k\varepsilon}\frac{\Gamma(2L+2)}{\Gamma(L+1-{\rm i}Z/(2k))}
\\
b_{1+}&=&C_1(2k)^{-L-1}{\rm e}^{{\rm i}k\pi\varepsilon/2}
{\rm e}^{{\rm i}\pi(3L+3)}{\rm e}^{-\frac{\pi Z}{4k}}
{\rm e}^{-k\varepsilon}\frac{\Gamma(2L+2)}{\Gamma(L+1-{\rm i}Z/(2k))}
\\
b_{1-}&=&C_1(2k)^{-L-1}{\rm e}^{-{\rm i}k\pi\varepsilon/2}
{\rm e}^{-\frac{\pi Z}{4k}}
{\rm e}^{-k\varepsilon}\frac{\Gamma(2L+2)}{\Gamma(L+1+{\rm i}Z/(2k))}
\\
a_{2+}&=&C_2(2k)^{L}{\rm e}^{-{\rm i}k\pi\varepsilon/2}
{\rm e}^{{\rm i}\pi 2L}{\rm e}^{-\frac{\pi Z}{4k}}
{\rm e}^{k\varepsilon}\frac{\Gamma(-2L)}{\Gamma(-L+{\rm i}Z/(2k))}
\\
a_{2-}&=&C_2(2k)^{L}{\rm e}^{{\rm i}k\pi\varepsilon/2}
{\rm e}^{-{\rm i}\pi L}{\rm e}^{-\frac{\pi Z}{4k}}
{\rm e}^{k\varepsilon}\frac{\Gamma(-2L)}{\Gamma(-L-{\rm i}Z/(2k))}
\\
b_{2+}&=&C_2(2k)^{L}{\rm e}^{{\rm i}k\pi\varepsilon/2}
{\rm e}^{-{\rm i}\pi 3L}{\rm e}^{-\frac{\pi Z}{4k}}
{\rm e}^{-k\varepsilon}\frac{\Gamma(-2L)}{\Gamma(-L-{\rm i}Z/(2k))}
\\
b_{2-}&=&C_2(2k)^{L}{\rm e}^{-{\rm i}k\pi\varepsilon/2}
{\rm e}^{-\frac{\pi Z}{4k}}
{\rm e}^{-k\varepsilon}\frac{\Gamma(-2L)}{\Gamma(-L+{\rm i}Z/(2k))}
\label{abs}
\end{eqnarray}
Note that as expected from the functional form of the solutions, 
the interchange of the indices $1\leftrightarrow 2$ corresponds to 
the interchange $L\leftrightarrow -L-1$. These coefficients are 
also related by the expressions 
\begin{equation}
b_{1+}=a_{1-}{\rm e}^{-2k\varepsilon}{\rm e}^{2{\rm i}\pi(L+1)}\hskip 2cm 
b_{1-}=a_{1+}{\rm e}^{-2k\varepsilon}{\rm e}^{-2{\rm i}\pi(L+1)}
\label{ab1}
\end{equation}
\begin{equation}
b_{2+}=a_{2-}{\rm e}^{-2k\varepsilon}{\rm e}^{2{\rm i}\pi(-L)}\hskip 2cm 
b_{2-}=a_{2+}{\rm e}^{-2k\varepsilon}{\rm e}^{2{\rm i}\pi L}
\label{ab2}
\end{equation}

The asymptotic expansion of the general wave function 
$\Phi(x(s)) = \alpha\psi_1(x(s)) + \beta\psi_2(x(s))$ is then the 
following:
\begin{equation}
\Phi(s\rightarrow -\infty)  \sim 
         (\alpha a_{1-}+\beta a_{2-}) 
	{\rm e}^{{\rm i}(k s -\frac{Z}{2k}\ln(-2ks))} 
	+ (\alpha b_{1-}+\beta b_{2-}) 
	 {\rm e}^{-{\rm i}(k s -\frac{Z}{2k}\ln(-2ks))}
\label{phiassn}
\end{equation}
\begin{equation}
\Phi(s\rightarrow \infty)  \sim 
        (\alpha a_{1+}+\beta a_{2+}) 
        {\rm e}^{{\rm i}(k s +\frac{Z}{2k}\ln(2ks))} 
	+ (\alpha b_{1+}+\beta b_{2+})
	b_{j+}  {\rm e}^{-{\rm i}(k s +\frac{Z}{2k}\ln(2ks))}\ . 
\label{phiassp}
\end{equation}
In the next step we may construct solutions that correspond to
an incoming wave from one direction in order to evaluate the
reflection and transmission coefficients \cite{fcjpdav}. 
Using the coefficients above the following results are obtained 
after some manipulations with gamma, trigonometric and exponential 
functions:
\begin{eqnarray}
T_{L\rightarrow R}(k)
&=& \frac{\rm i}{2\pi}
{\rm e}^{-{\rm i}\pi k\varepsilon}{\rm e}^{\frac{\pi Z}{2k}}
\Gamma(-L-{\rm i}Z/(2k))\Gamma(L+1-{\rm i}Z/(2k))
\label{tlr}
\\
R_{L\rightarrow R}(k)
&=& 
T_{L\rightarrow R}(k)
{\rm e}^{-2 k\varepsilon}\left(-{\rm e}^{-\frac{\pi Z}{k}}+2\cos(2\pi L)\right)
\label{rlr}
\\
T_{R\rightarrow L}(k)
&=& 
-T_{L\rightarrow R}(k)
\label{trl}
\\
R_{R\rightarrow L}(k)
&=& 
T_{R\rightarrow L}(k)
{\rm e}^{2 k\varepsilon}{\rm e}^{\frac{-\pi Z}{k}}
\label{rrl}
\end{eqnarray}

It is seen that as expected, the bound-state energies emerge from the 
poles of $T_{L\rightarrow R}(k)$ (\ref{tlr}), i.e. when the arguments 
of the gamma functions are set to the $-n$ non-positive integer. 

Equation (\ref{trl}) is also in accordance with the results of 
scattering on other ${\cal PT}$-symmetric potentials in that the 
transmission coefficient does not show handedness effect, except 
for a trivial factor of $-1$, while the reflection coefficients 
clearly demonstrate handedness \cite{ahmedxx,fcjpdav}. This means 
that similarly to other examples, waves arriving from the 
asymptotically absorptive side scatter differently from waves 
arriving from the asymptotically emissive one. 

To exemplify the results we present in Fig. \ref{Potential} the 
potential as the function 
of $s$ for some parameter, together with the transmission and 
reflection coefficients in Fig. \ref{TR} for a series of parameters. 
\begin{figure}[ht]
\begin{center}
\subfloat[ $Z=1,L=3.75,\varepsilon=0.005$ ]{
\includegraphics[ width =0.45\textwidth ]{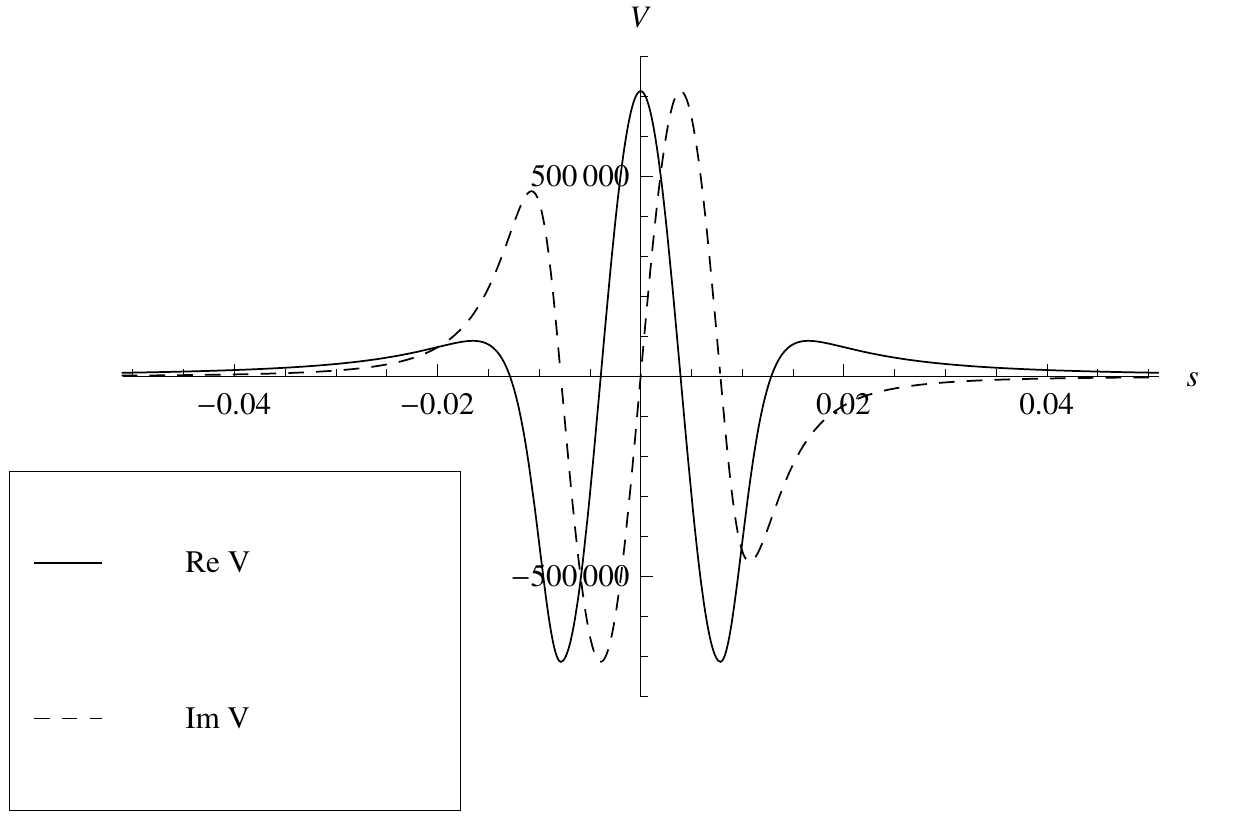}}
\subfloat[ $Z=1,L=3.53,\varepsilon=0.005$ ]{
\includegraphics[ width =0.45\textwidth ]{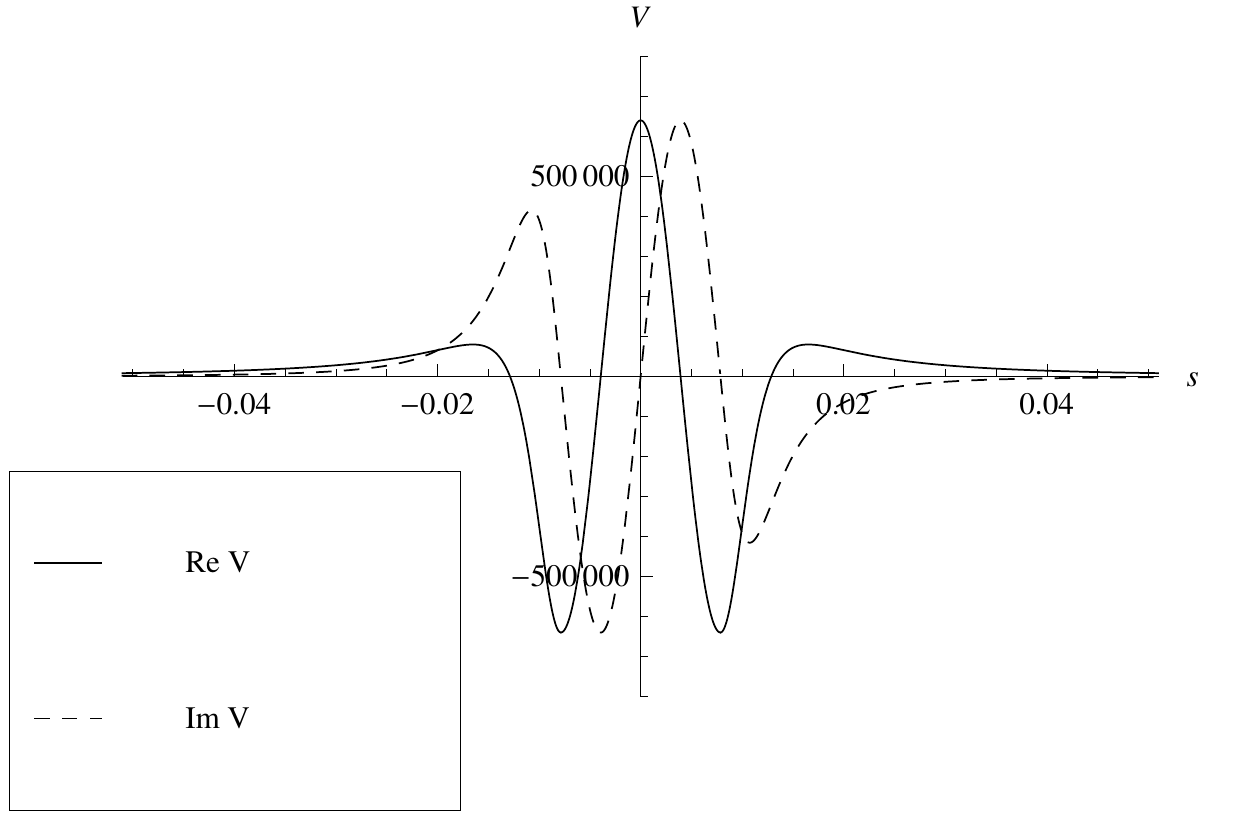}}\\
\subfloat[ $Z=1,L=3.10,\varepsilon=0.005$ ]{
\includegraphics [width =0.45\textwidth ]{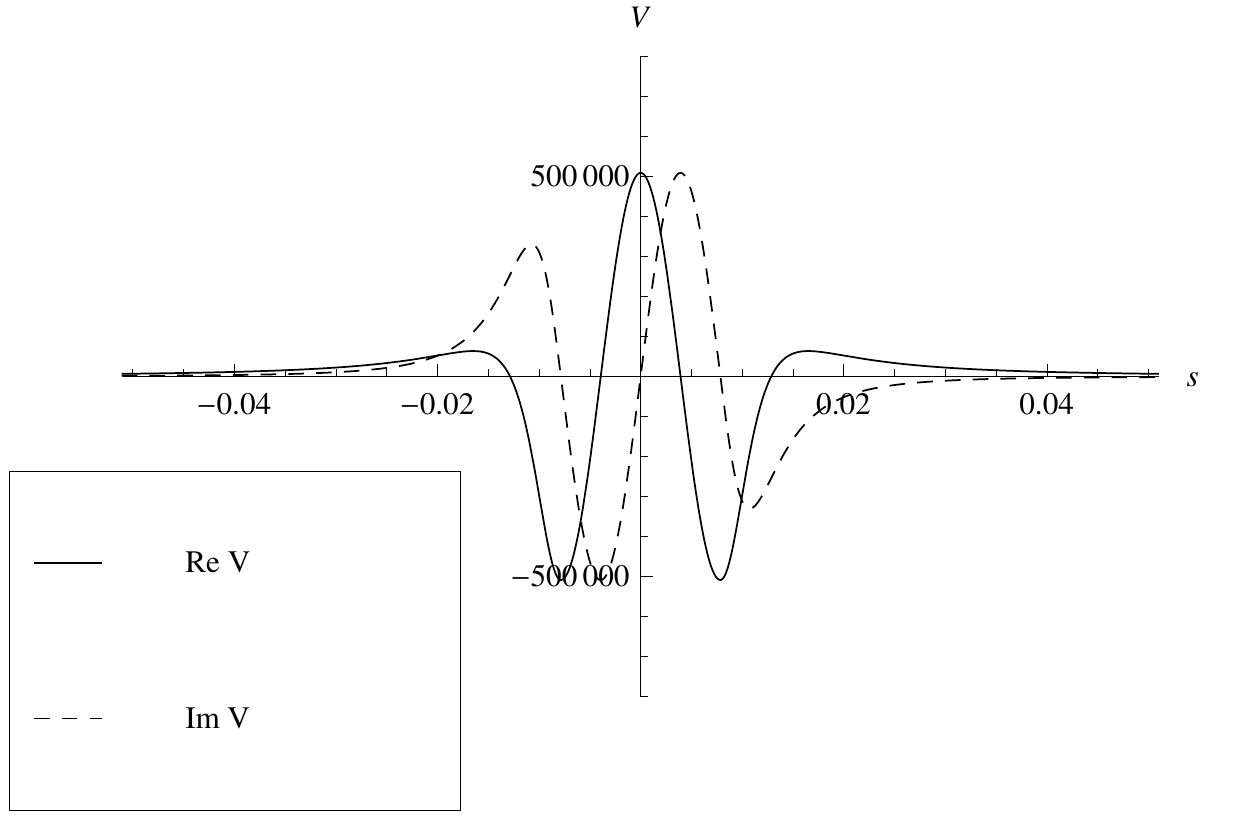}}
\subfloat[ $Z=1,L=3.01,\varepsilon=0.005$ ]{
\includegraphics[width =0.45\textwidth ]{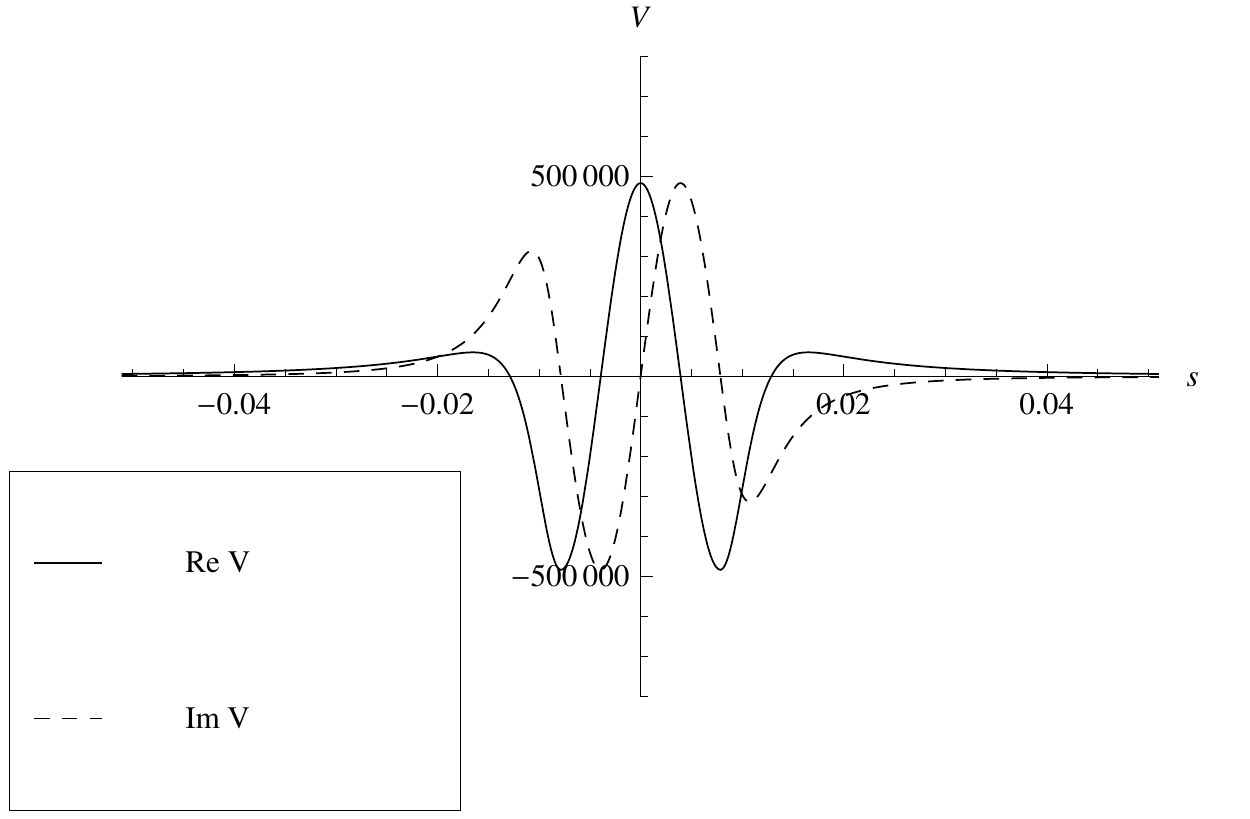}}
\caption {The ${\cal PT}$-symmetric Coulomb potential plotted as the 
function of $s$ parameter for $Z=1$, $\varepsilon=0.005$ and some 
values of $L$. }
\label{Potential}
\end{center}
\end{figure}
As expected, the real and the imaginary potential components are 
even and odd functions of $s$, furthermore, the real component 
decays quicker than the imaginary one as it should based on (\ref{SEor}). 
The role of the $\varepsilon$ parameter is essentially that of a 
length scale. $L$ and $Z$ set the scale of the real and 
imaginary potential component, respectively. 

\begin{figure}[ht]
\begin{center}
\subfloat[ $Z=1,L=3.75,\varepsilon=0.005$ ]{
\includegraphics[ width =0.45\textwidth ]{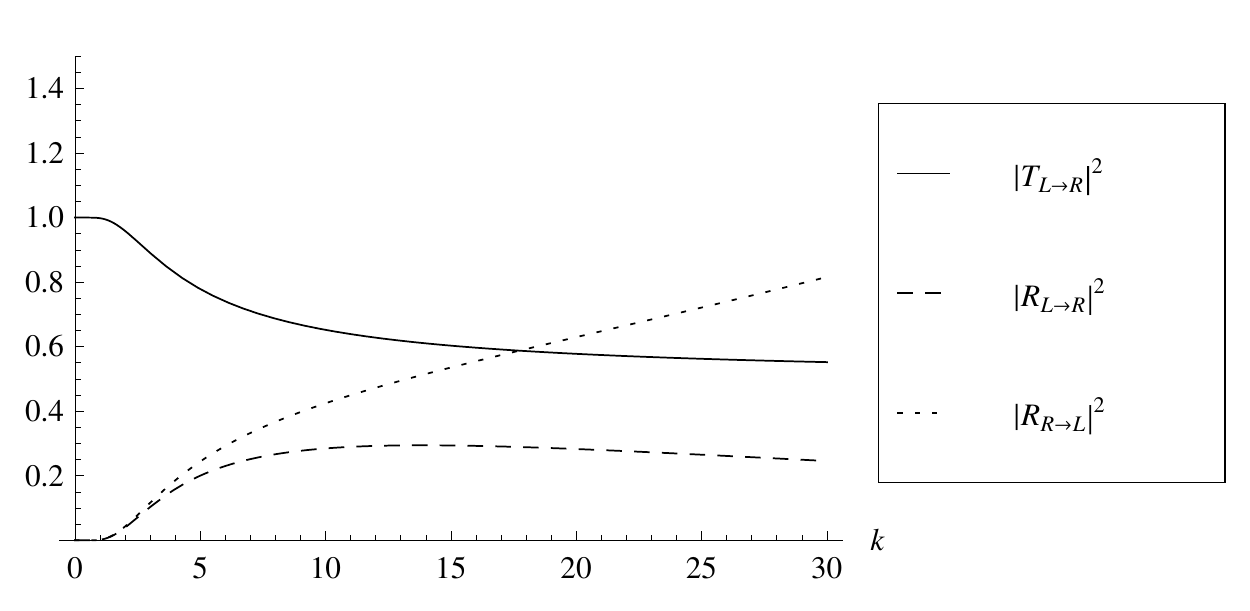}}
\subfloat[ $Z=1,L=3.53,\varepsilon=0.005$ ]{
\includegraphics[ width =0.45\textwidth ]{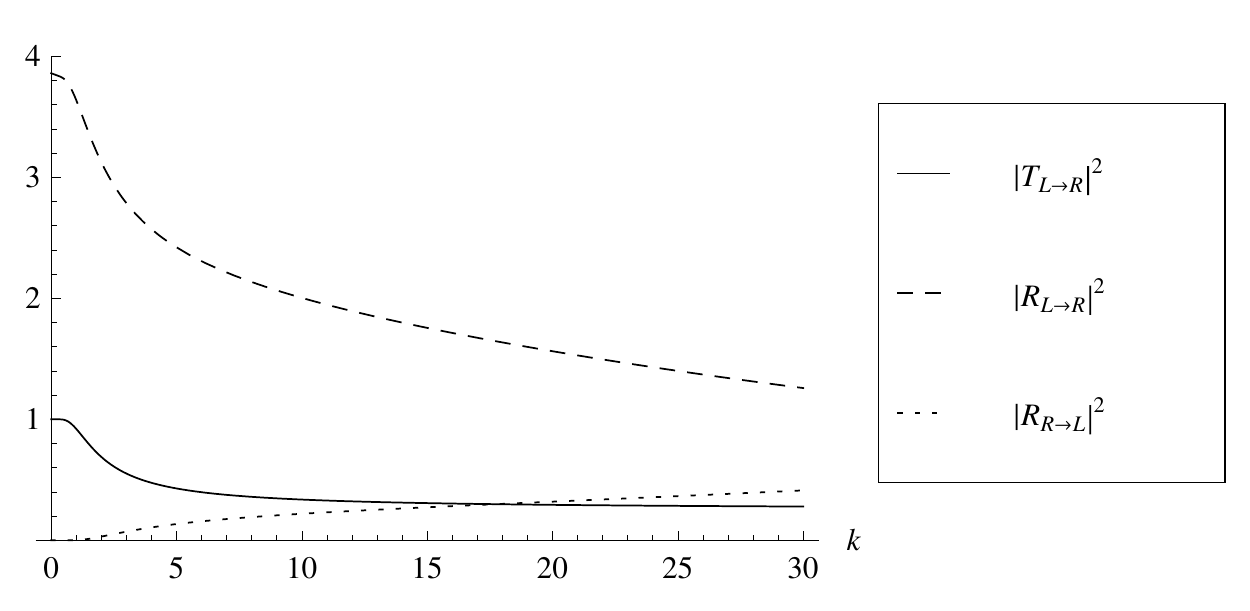}}\\
\subfloat[ $Z=1,L=3.10,\varepsilon=0.005$ ]{
\includegraphics [width =0.45\textwidth ]{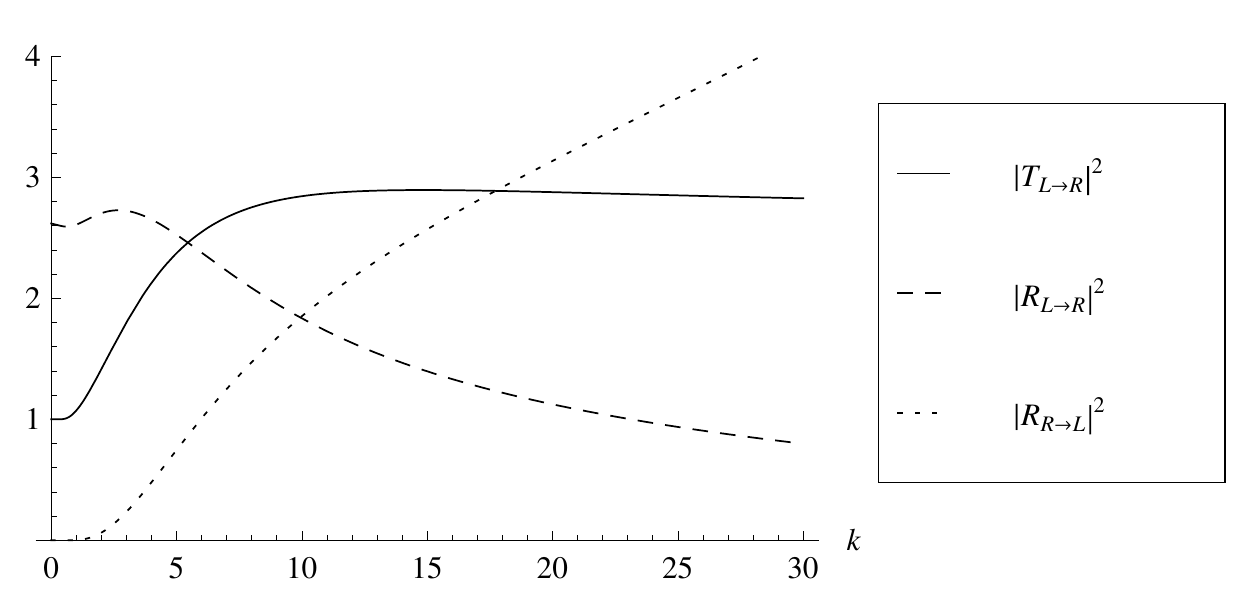}}
\subfloat[ $Z=1,L=3.01,\varepsilon=0.005$ ]{
\includegraphics[ width =0.45\textwidth ]{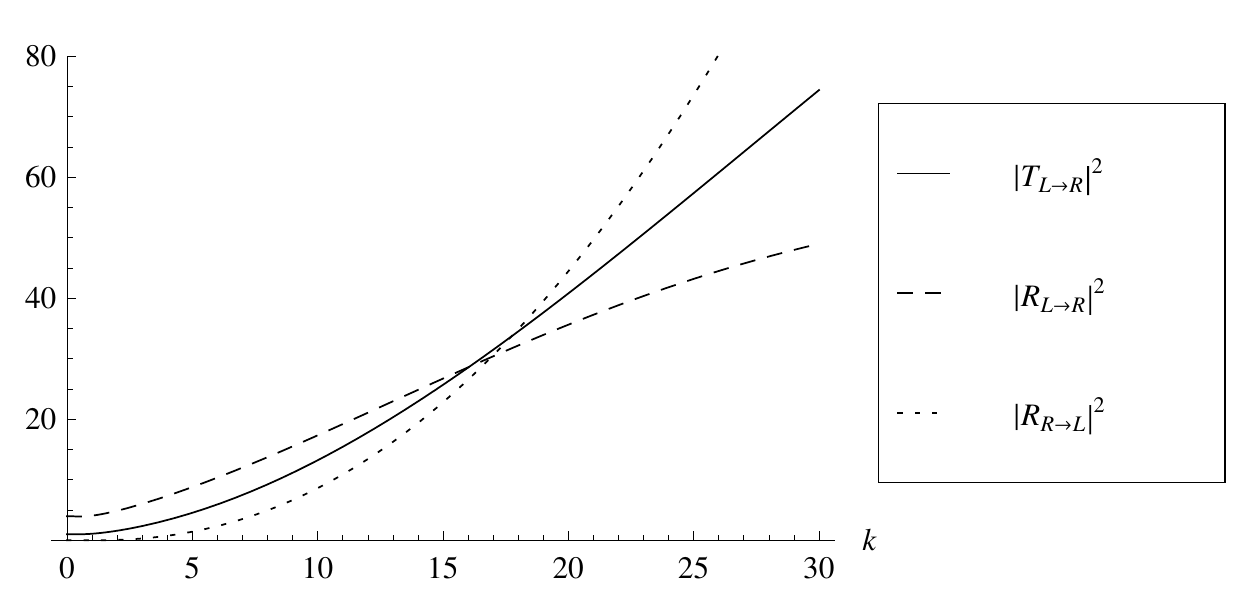}}
\caption {Transmission and reflection coefficients corresponding to 
the parameters used in Fig. \ref{Potential}. }
\label{TR}
\end{center}
\end{figure}

It is seen that the large-$k$ behaviour of $T_{L\rightarrow R}(k)$ 
is relatively smooth, while that of the reflection coefficients 
is dominated by the $\exp(\pm 2k\varepsilon)$ factors. An exception 
for this behaviour is seen only near integer values of $L$, which 
are excluded from the present discussion. The $Z$ ``charge'' parameter 
does not influence the results in an essential way.

\section{Summary}
\label{sum}

Scattering on the ${\cal PT}$-symmetric Coulomb potential was 
discussed on a trajectory of the complex $x$ plane. This trajectory 
was a U-shaped curve circumventing the origin from below, guaranteeing 
the asymptotical regularity of bound states. From the topological 
point of view it is similar to the parabolic path obtained from 
the application of the harmonic oscillator--Coulomb mapping of 
bound states to the ${\cal PT}$-symmetric setting \cite{pla00}. 
It is also reminiscent to the trajectories obtained by similar 
regularity arguments for the power-type potentials appearing in 
the first publications of ${\cal PT}$-symmetric quantum mechanics 
\cite{bb98}: there the allowed wedges for the solutions were located in 
the lower half of the complex $x$ plane, while here the allowed domain 
is the upper half, corresponding formally to the inverted power 
$({\rm i}x)^{-1}$ expression of the Coulomb potential. In 
addition to the Coulombic potential term a centrifugal-like 
term $L(L+1)x^{-2}$ was also considered. Here integer values of 
$L$ were excluded, because in that case the linearly 
independent solutions had to be defined in another way. 

Parametrizing the trajectory as $x(s)$ in terms of the real parameter 
$s\in(-\infty,\infty)$ allowed the expression of the asymptotic 
solutions in a form familiar from the discussion of the real 
Coulomb potential. When parametrized in terms of $s$, the real 
and the imaginary potential components vanished as $s^{-2}$ and 
$s^{-1}$, i.e. the imaginary potential component dominates the 
problem asymptotically. 

The transmission and reflection coefficients were determined, 
from the asymptotic solutions, with special attention to the continuity 
of the solutions in the complex plane. It was found that the 
transmission coefficients for waves arriving from the two directions 
differ only in a trivial factor of $-1$, while the reflection 
coefficients exhibit manifest handedness. This is similar to 
other examples for scattering in ${\cal PT}$-symmetric potentials, 
which, however, were defined on the real $x$ axis or its trivial 
shifted version $x-{\rm i}c$. The $\varepsilon$ parameter appearing in the definition of the 
U-shaped curve plays a role similar to the $c$ parameter applied 
in Refs. \cite{jpa01,jpa02b} to shift the trajectory off the real $x$ axis. 
Although it changes the potential shape, it does not 
influence the energy spectrum.

The results showed strong dependence on the $\varepsilon$ parameter 
that sets the distance of the U-shaped trajectory from the positive 
imaginary axis, while the $Z$ ``charge'' parameter did not influence 
the results in an essential way. It essentially sets the relative 
weight of the imaginary and real potential components. 
Dependence on the $L$ parameter was 
also significant in that the transmission and reflection coefficients 
showed rapid variations close to the forbidden integer values of $L$. 

As another aspect similar to other scattering problems, both Hermitian 
and ${\cal PT}$-symmetric, the two sets of bound-state energy 
eigenvalues described in Ref. \cite{pla00} could be recovered from the 
poles of the transmission coefficients. Furthermore, the reparametrization 
of the problem in terms of the $s$ variable also allowed to recover 
the energy spectrum with the correct sign: $E<0$ for bound states  
and $E>0$ for scattering states.

\section*{Acknowledgements}
M. Z. appreciates the support by the GA\v{C}R grant No.
202/07/1307, P. S. appreciates the support by the GA\v{C}R grant No.
202/07/1307 and CTU grant No. CTU0910114, G. L. acknowledges the support by the OTKA grant
No. T49646. Thanks of M. Z. and P. S. for the hospitality of the Institute
of Nuclear Research of the Hungarian Academy of Sciences in
Debrecen should be added, together with the acknowledgement of the
support of M. Z. and P. S. by the ``Doppler Institute" project No.
LC06002 (M\v{S}MT) and by the Institutional Research Plan
AV0Z10480505. Our collaboration project is a part of the bilateral
three-year exchange program between the Hungarian and Czech
Academies of Sciences.

\end{document}